\documentclass{PoS}
\usepackage[export]{adjustbox}

\title{Energy dependence of hadron spectra and multiplicities in p+p interactions}

\ShortTitle{Energy dependence of hadron spectra and multiplicities in p+p interactions}

\author{\speaker{Szymon Pu\l awski}\thanks{A footnote may follow.}\\
		for the NA61/SHINE Collaboration\\
        Institute of Physics, University of Silesia, Katowice, Poland\\
        E-mail: \email{s.pulawski@cern.ch}}

\abstract{The NA61/SHINE experiment at the CERN SPS aims to discover the critical point of strongly
interacting matter and study the properties of the onset of deconfinement. In order to reach these goals measurements of hadron production properties are performed in nucleus-nucleus,
proton-proton and proton-nucleus interactions as a function of collision energy and size of the
colliding nuclei.

Inclusive spectra of identified hadrons in p+p interactions
at the SPS energies are presented as a function of transverse momentum, transverse mass and rapidity. 
The results are compared with the world data and theoretical models.}

\FullConference{9th International Workshop on Critical Point and Onset of Deconfinement - CPOD2014,\\
		17-21 November 2014\\
		ZiF (Center of Interdisciplinary Research), University of Bielefeld, Germany}

\begin{document}

\section{Experimental data and analysis methods}
$\>$ This paper presents inclusive spectra of identified
hadrons produced in inelastic p+p interactions at 20, 31, 40, 80, 158~GeV/c. 
Three analysis methods were used:
\begin {itemize}
	\item For $\pi^-$ spectra the $h^-$ method  \cite{Abgrall:2013qoa}, which is based
on the fact that the majority of negatively charged particles are $\pi^{-}$ mesons. 
The contribution of other particles is subtracted using calculations based 
on the EPOS model \cite{Pierog:2009zt},
	\item for $\pi^{+}$, $\pi^{-}$, $K^{+}$, $K^{-}$ and $p$, $\bar{p}$ the $dE/dx$ method, which uses information on particle energy loss $(dE/dx)$ in the TPC gas to identify particles,
 	\item for $\pi^{+}$, $\pi^{-}$, $K^{+}$, $K^{-}$ and $p$, $\bar{p}$ the $tof-dE/dx$ method, which in addition to the $dE/dx$ information uses time of flight $(tof)$ measurements. Combined $tof-dE/dx$ analysis provides excellent separation of different hadron species close to the mid-rapidity region.
\end {itemize}

Results were corrected for trigger and event selection biases,
contribution  of non-target interactions as well as detector inefficiencies and feed-down from weak decays and secondary
interactions. 

\section{Results on single-particle spectra}

\begin{figure}[!ht]
	\begin{center}
	\includegraphics[width=0.3\textwidth]{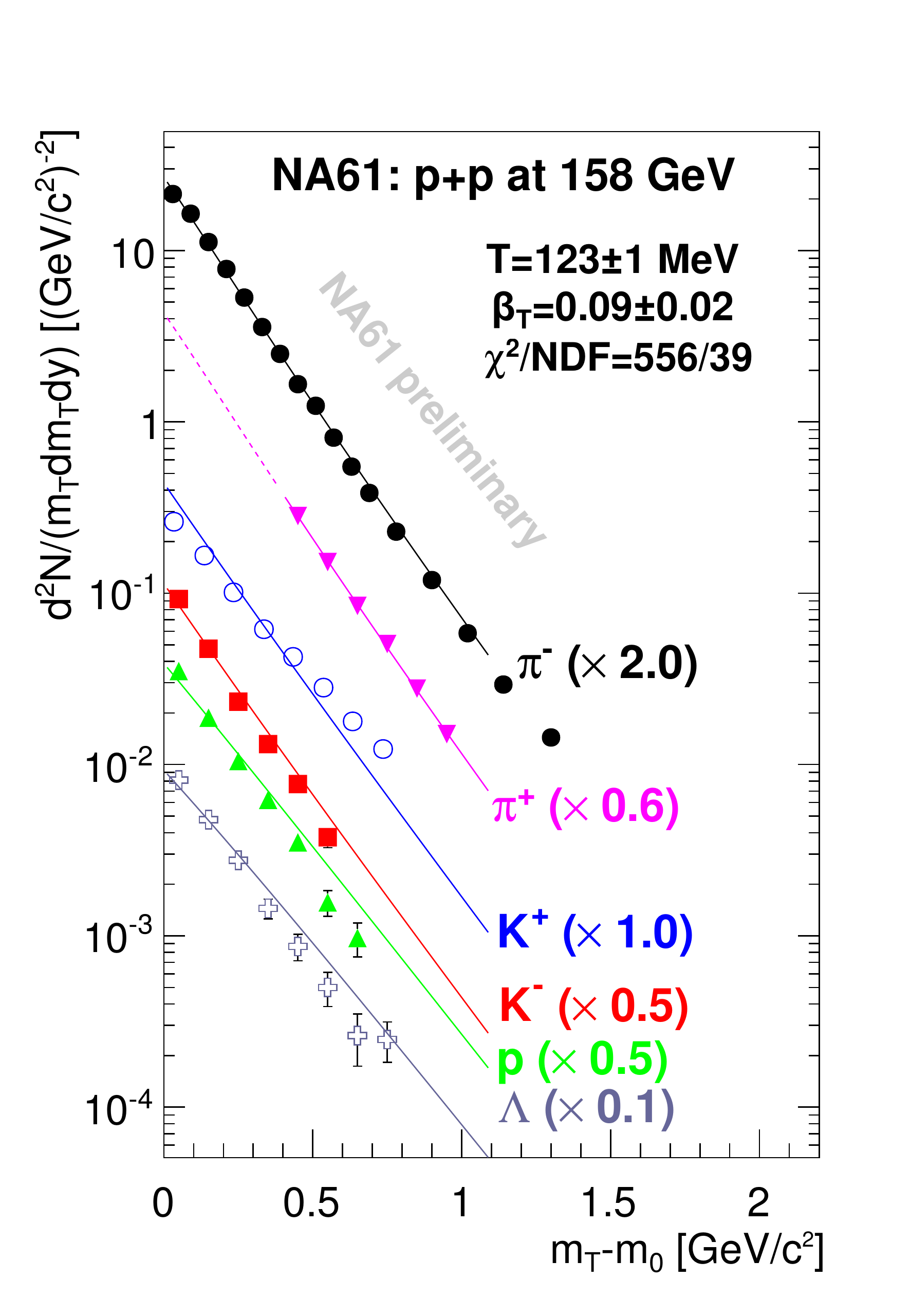}
   	 \includegraphics[width=0.3\textwidth]{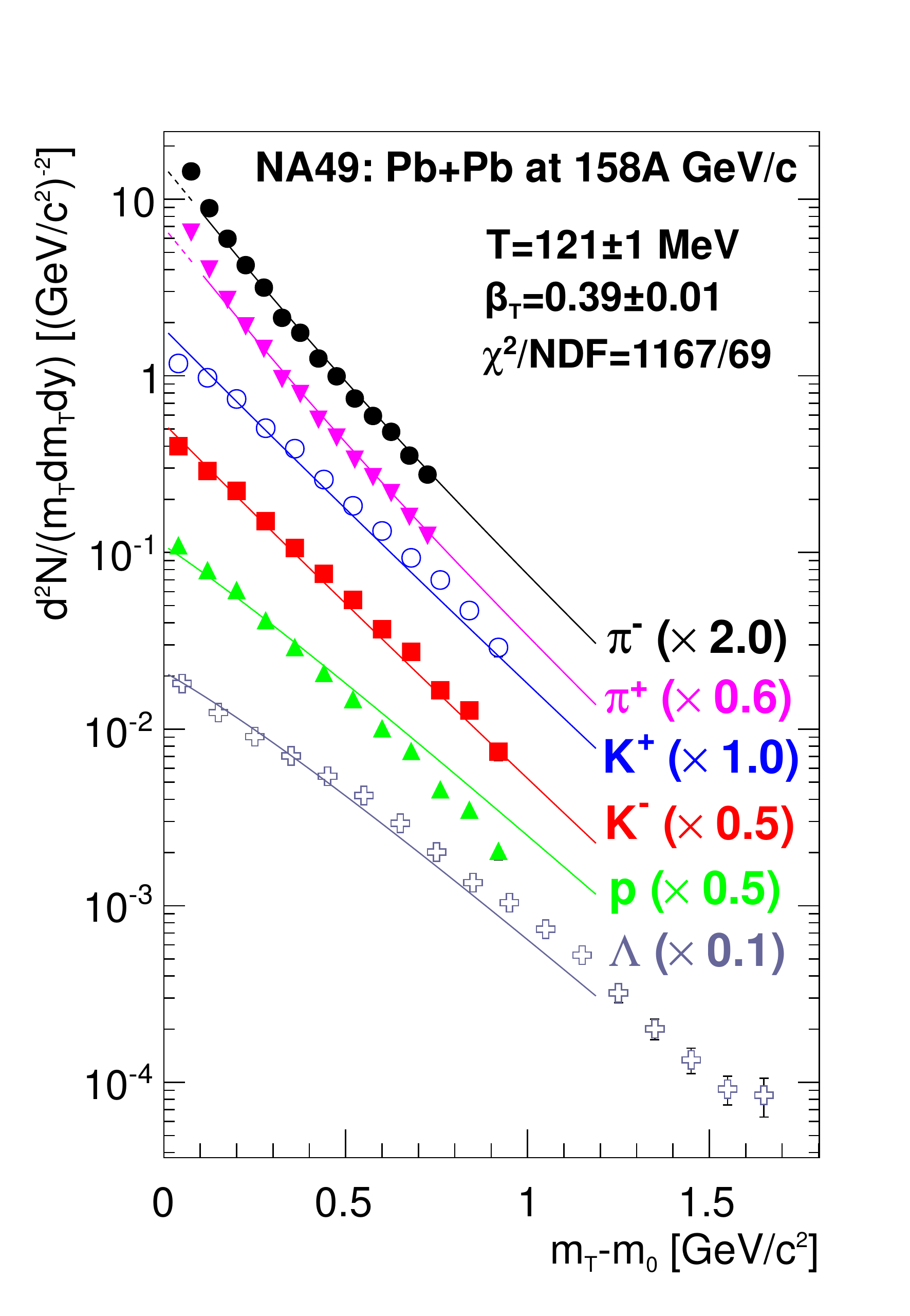}
	\end{center}
	\caption{Transverse mass spectra at mid-rapidity measured by the NA61/SHINE experiment in inelastic p+p interactions (left) and corresponding results of NA49 results from central Pb+Pb collisions at 158$A$~GeV (right).}
	\label{fig:blast}
\end{figure}

Spectra of transverse mass $m_T$ of negatively and positively charged pions, kaons, protons and $\Lambda$ hyperons produced in inelastic p+p interactions at 158~GeV/c at mid-rapidity are presented in Fig.~\ref{fig:blast} (left). The corresponding data for central Pb+Pb collisions measured by NA49 \cite{Alt:2006dk,Alt:2007aa,Afanasiev:2002mx} are shown in the right panel of Fig. \ref{fig:blast} (right). These data sets were fitted using the simplified blast wave model parameterization~\cite{Schnedermann:1993ws}:
$\frac{dN_{i}}{m_{T}dm_{T}dy} = A_{i}m_{T}K_{1}(\frac{m_{T}cosh\rho}{T})I_{0}(\frac{p_{T}sinh\rho}{T})$.
The transverse flow velocity  $\beta_{T}$ was calculated from the equation 
$\rho = tanh^{-1}\beta_{T}$. 

The value of $\beta_{T}$ is much higher in central Pb+Pb than in p+p reactions.
Transverse mass spectra are approximately exponential in p+p interactions.  The exponential dependence is modified by the transverse flow in central (7\%) Pb+Pb collisions. Transverse collective flow in Pb+Pb can explain the significant spectra 
shape difference between p+p and central Pb+Pb reactions.

\begin{figure}[!ht]
	\begin{center}
	\includegraphics[width=0.32\textwidth]{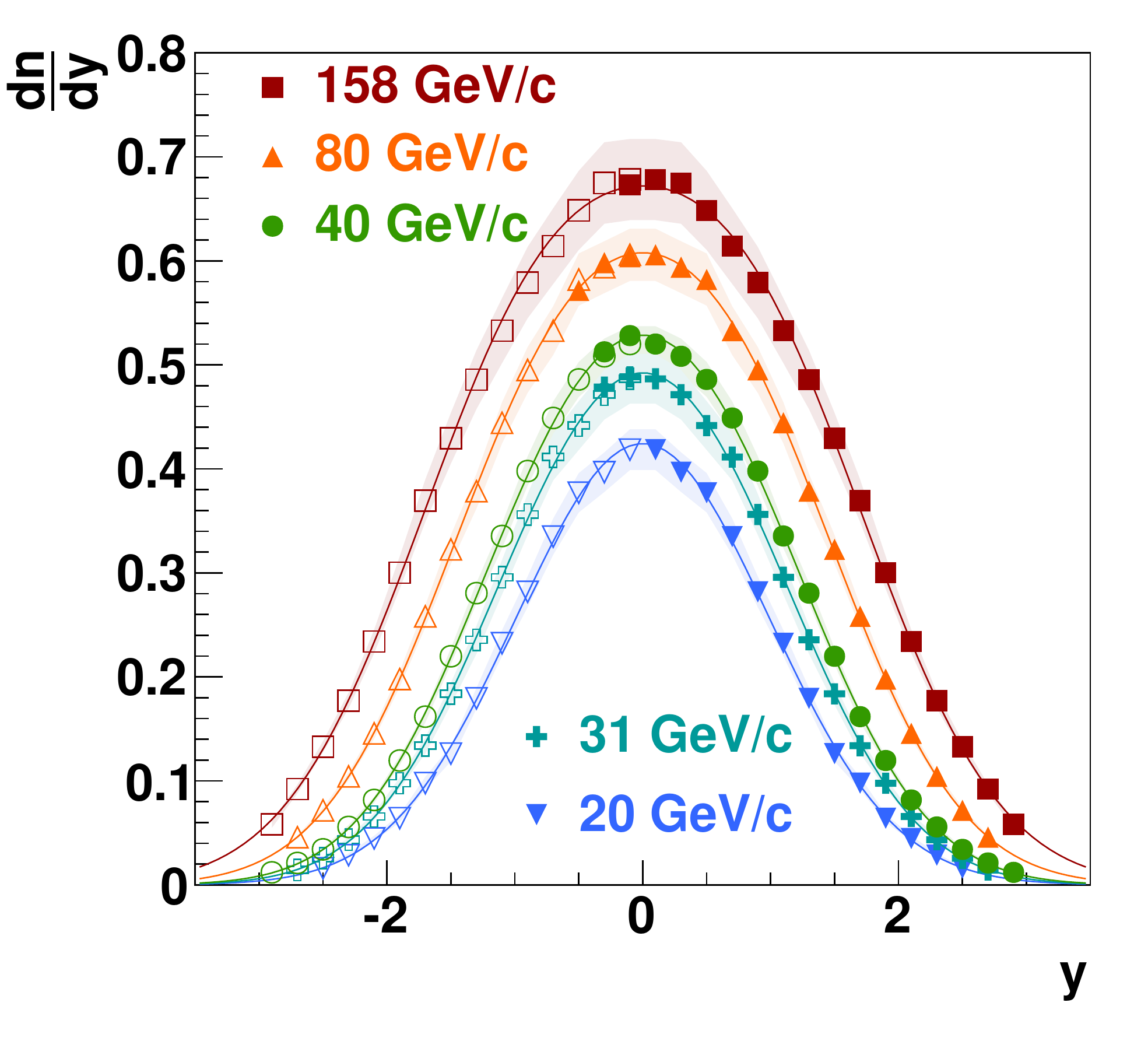}
	\includegraphics[width=0.32\textwidth]{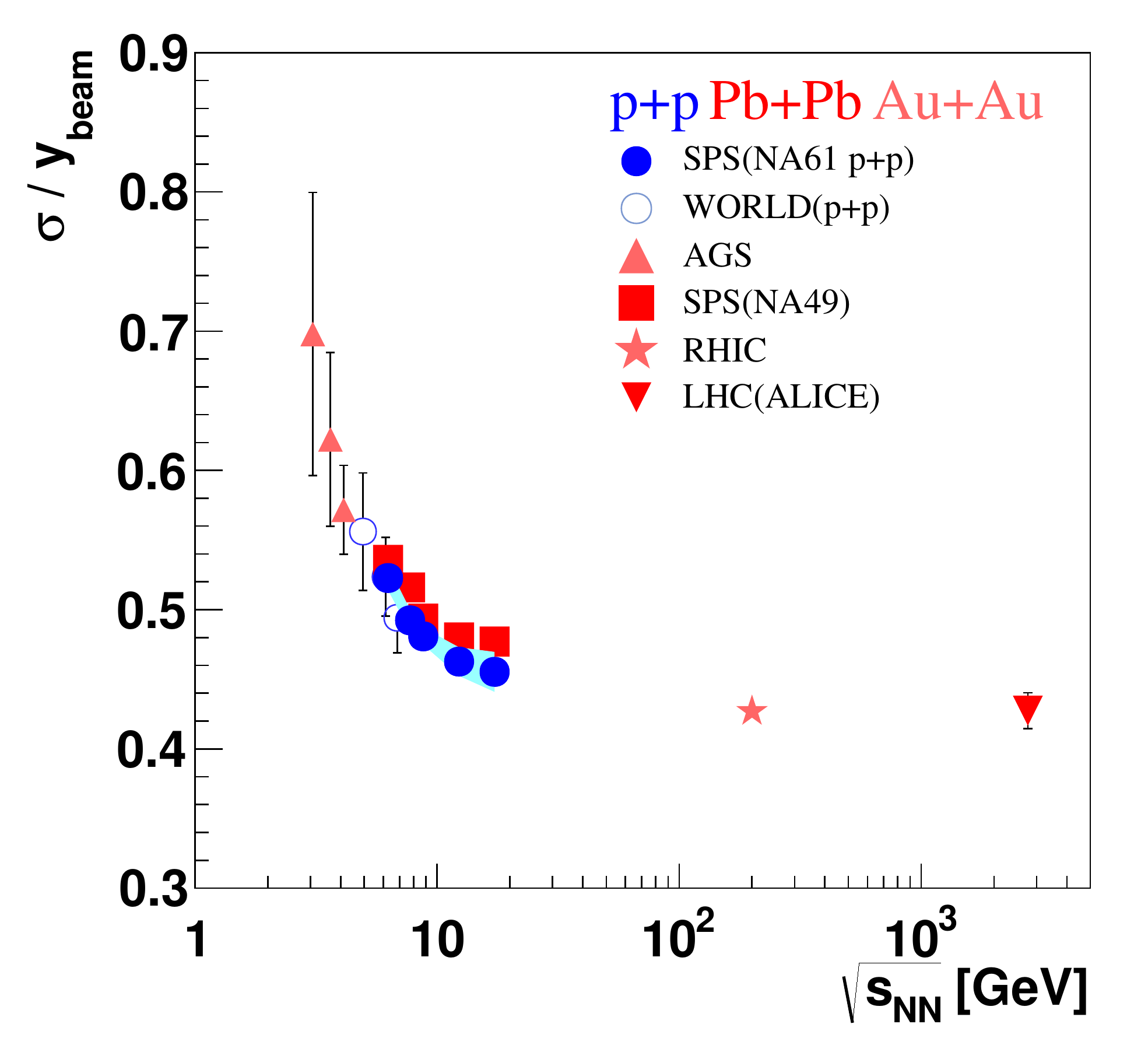}
	\includegraphics[width=0.32\textwidth]{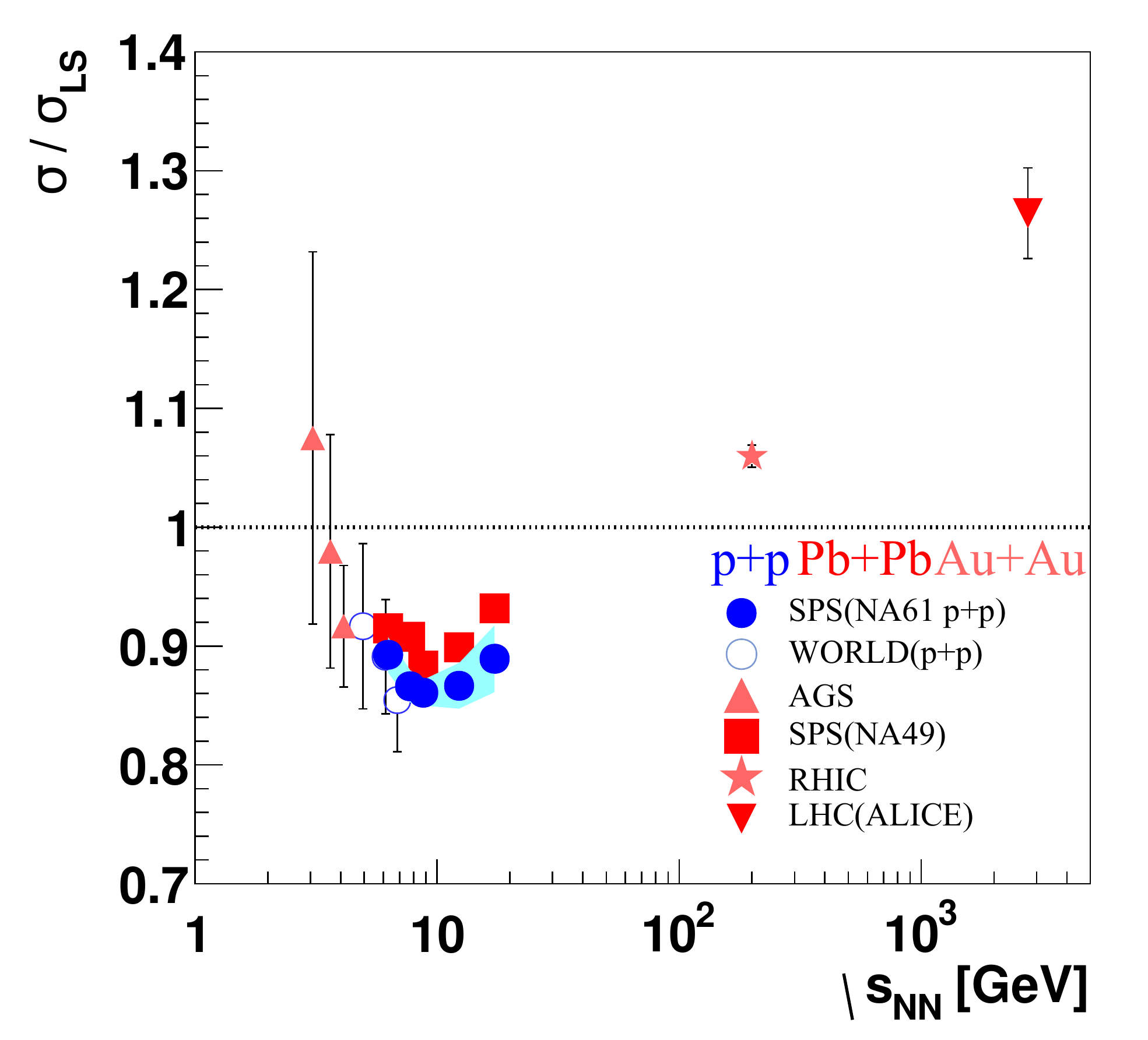}
	\end{center}
	\caption{$\pi^{-}$ rapidity distributions measured in inelastic p+p collisions at 20, 31, 40, 80 and 158 GeV/c (left). Dependence of the scaled width of the rapidity distribution on collision energy (center and right). World data are taken from Refs. \cite{Klay:2003zf,Abbas:2013bpa,Adler:2003cb}. Results for the width of rapidity distributions are not corrected for isospin effects.}
	\label{fig:pimirapidity}
\end{figure}

Rapidity distributions of $\pi^{-}$ produced in inelastic p+p collisions~\cite{Abgrall:2013qoa} and energy dependence of the width $\sigma$ of the rapidity distribution divided by beam rapidity ($y_{beam}$) or by the hydrodynamic model prediction ($\sigma_{LS}$)~\cite{Landau:1953gs,Shuryak:1974zc} are presented in Fig.~\ref{fig:pimirapidity}. The shape of the rapidity distribution is approximately Gaussian, however the best fit was obtained by a sum of two Gauss distributions. The width of the $\pi^{-}$ rapidity distribution divided by beam rapidity decreases with the collision energy. $\sigma/\sigma_{LS}$ and $\sigma/y_{beam}$ are smaller in p+p than in Pb+Pb interactions. No significant difference of the energy dependence is observed for the width of the $\pi^{-}$ rapidity distribution in p+p and Pb+Pb interactions. 

\begin{figure}[!ht]
	\begin{center}
	\includegraphics[width=0.4\textwidth]{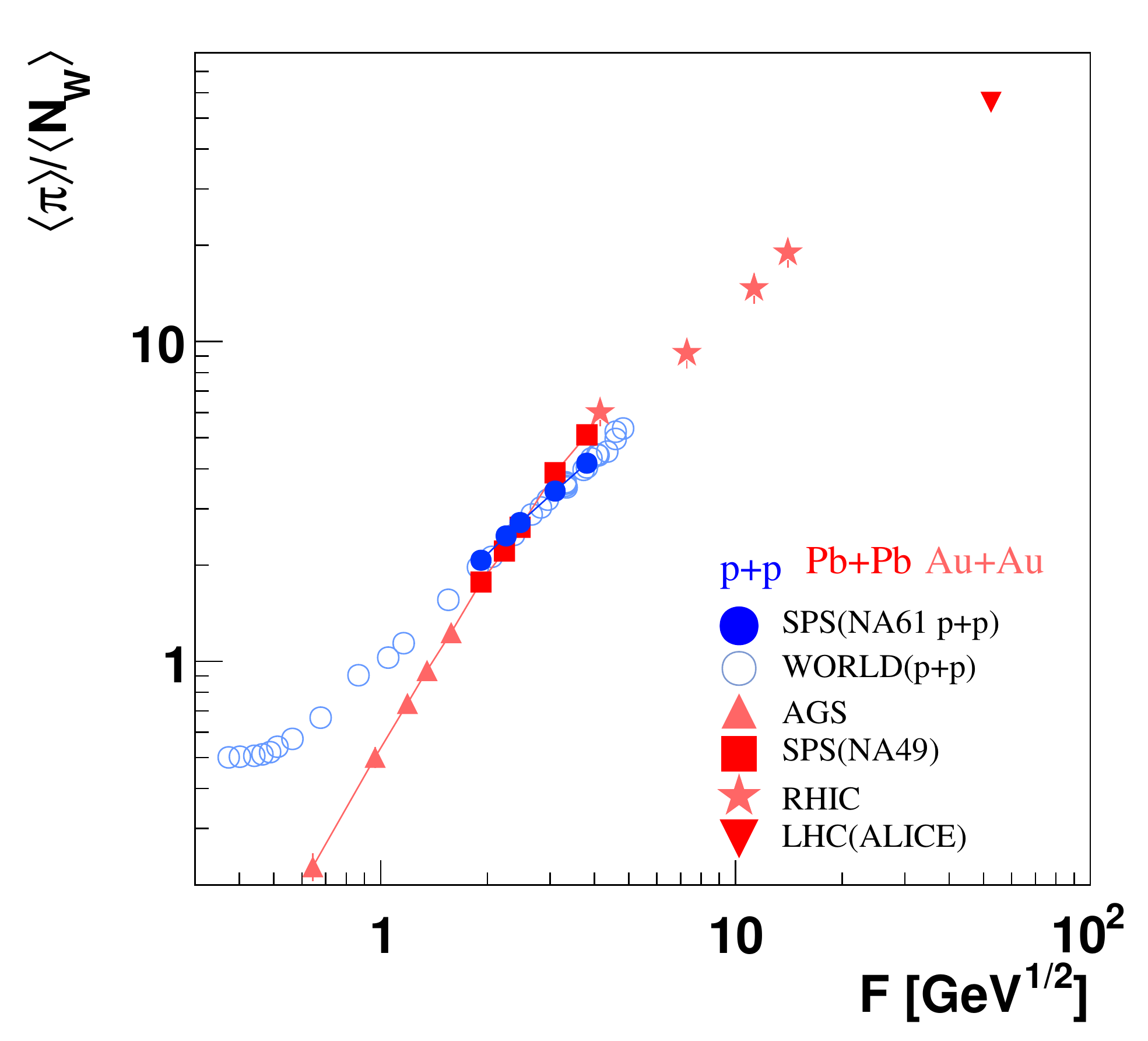}
	\includegraphics[width=0.4\textwidth]{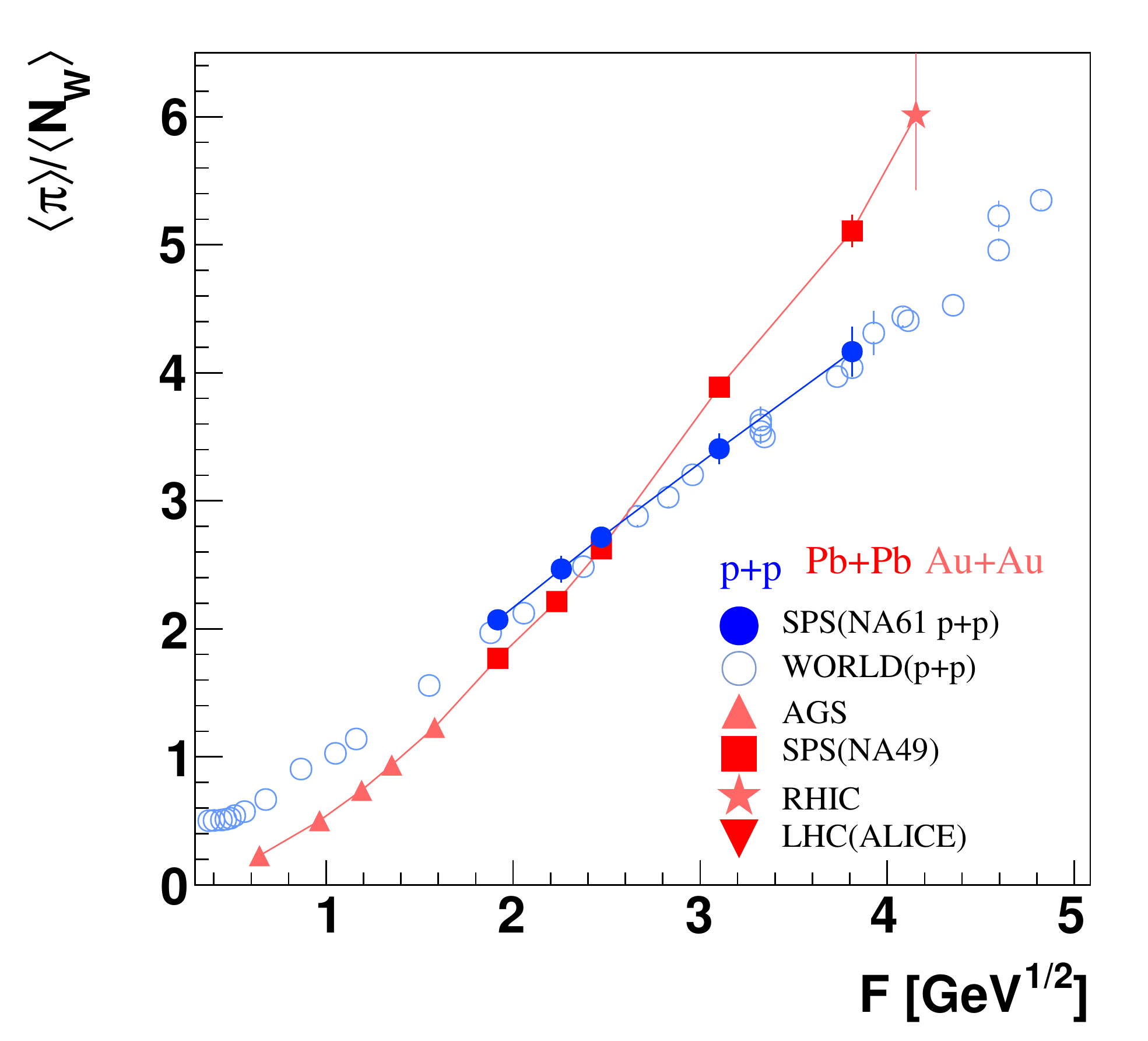}
	\end{center}
	\caption{Energy dependence of $\pi$ multiplicity as a function of collision energy. ALICE points were derived from Refs. \cite{Abbas:2013bpa,Abelev:2012wca}.}
	\label{fig:pimimultip}
\end{figure}

The results presented in this paper are an important step towards the study of the system size dependence of the signals of the onset of deconfinement observed in central Pb+Pb collisions, the ''kink'', ''horn'' and ''step'' \cite{Alt:2007aa, Gazdzicki:2010iv}. The $\pi$ multiplicity at SPS energies shown in Fig.~\ref{fig:pimimultip} increases faster in central Pb+Pb than in inelastic p+p collisions (kink) and the two dependencies cross each other at about 40\textit{A} GeV/c.

Preliminary NA61/SHINE measurements allow to significantly improve the world data on the inverse slope parameter $T$ of transverse mass spectra of kaons \cite{Kliemant:2003sa} as well as the ratio of $K^+$ and $\pi^+$ yields at mid-rapidity~\cite{Gazdzicki:1995zs,Gazdzicki:1996pk}.

\begin{figure}[!ht]
	\begin{center}
	\includegraphics[width=0.45\textwidth]{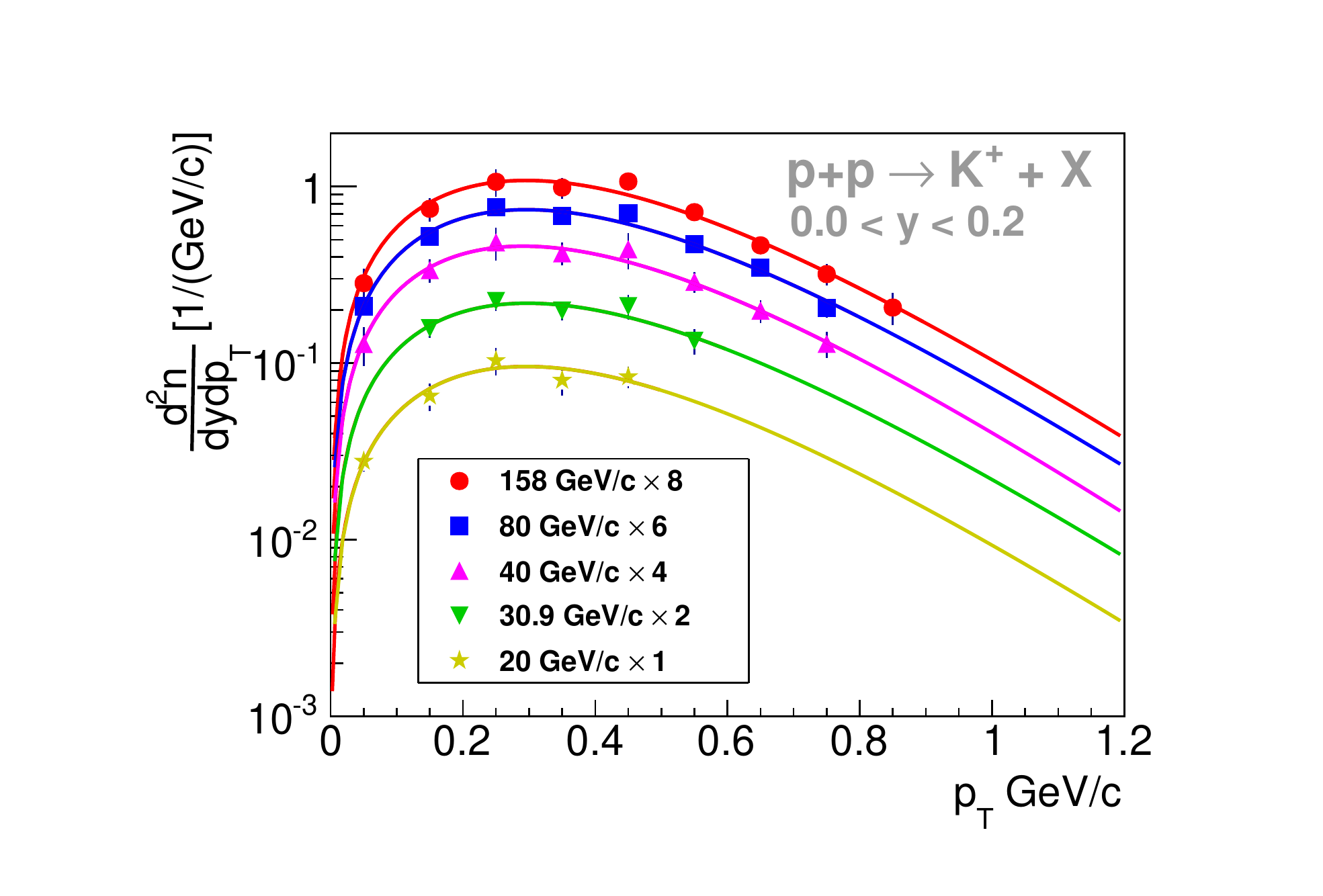}
	\includegraphics[width=0.45\textwidth]{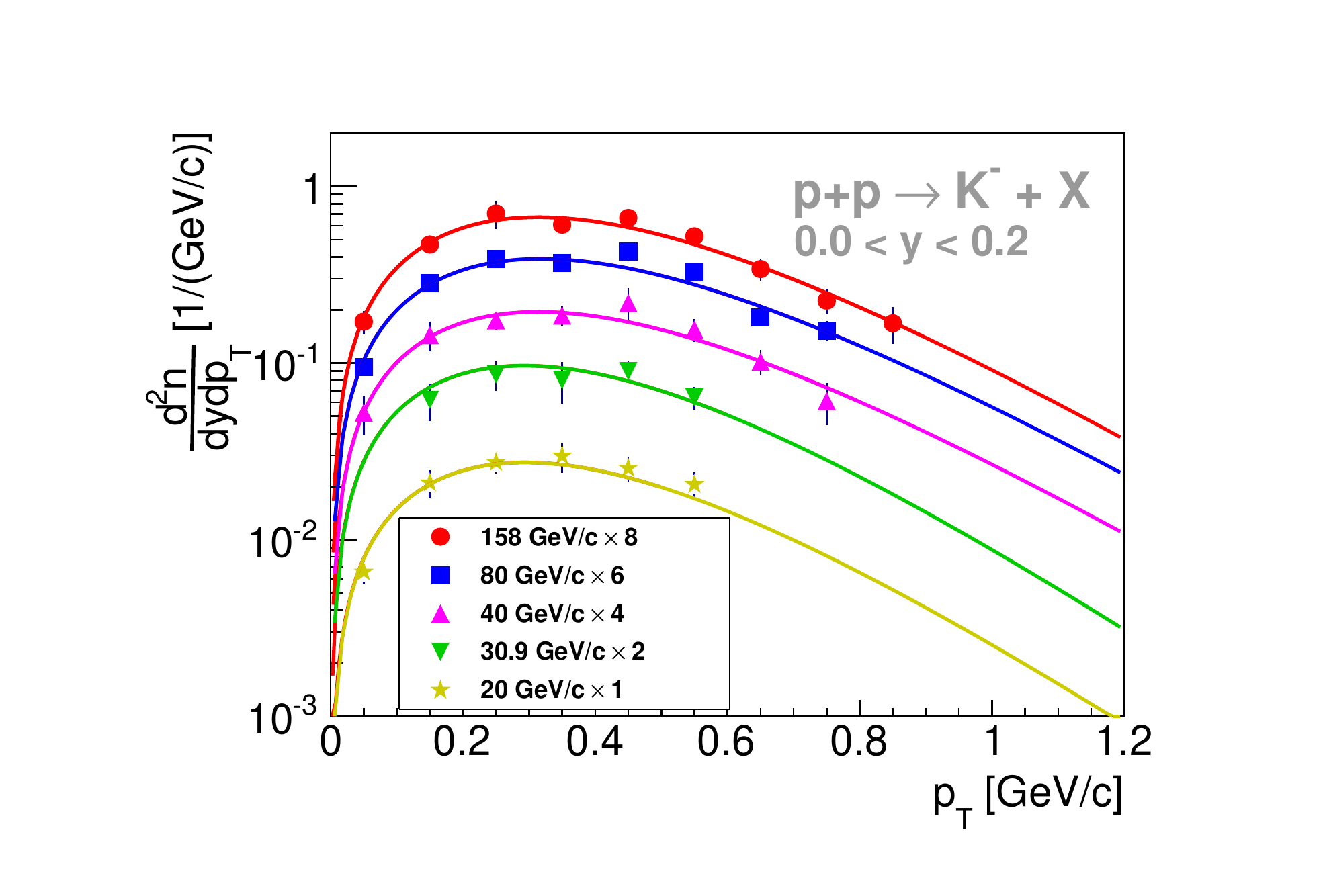}
	\end{center}
	\caption{Transverse momentum spectra of $K^{+}$ and $K^{-}$ mesons produced at $y \approx 0$ in inelastic p+p interactions measured by NA61/SHINE.}
	\label{fig:kaonpt}
\end{figure}

The m$_T$ spectra were parametrized by the exponential function:
\begin{equation}
%\frac{d^2n}{dp_Tdy}(p_T)_{y}=Ap_Te^{-\frac{\sqrt{p_T^2+m_K^2}}{T}},
 \frac{d^{2}n}{dp_{T}dy}=\frac{S~p_{T}}{T^{2}+m_{K}T}\exp\left(-\frac{\sqrt{p^{2}_{T}+
m^{2}_{K}}-m_{K}}{T}\right),
\label{eq:inverse}
\end{equation}
where $S$ and $T$ are the yield integral and  the inverse slope parameter, respectively.
The $S$ and $T$ parameters were fitted to spectra
of $K^{+}$ and $K^{-}$ mesons at mid-rapidity as shown in Fig.~\ref{fig:kaonpt}.

\begin{figure}[!ht]
	\begin{center}
	\includegraphics[width=0.48\textwidth]{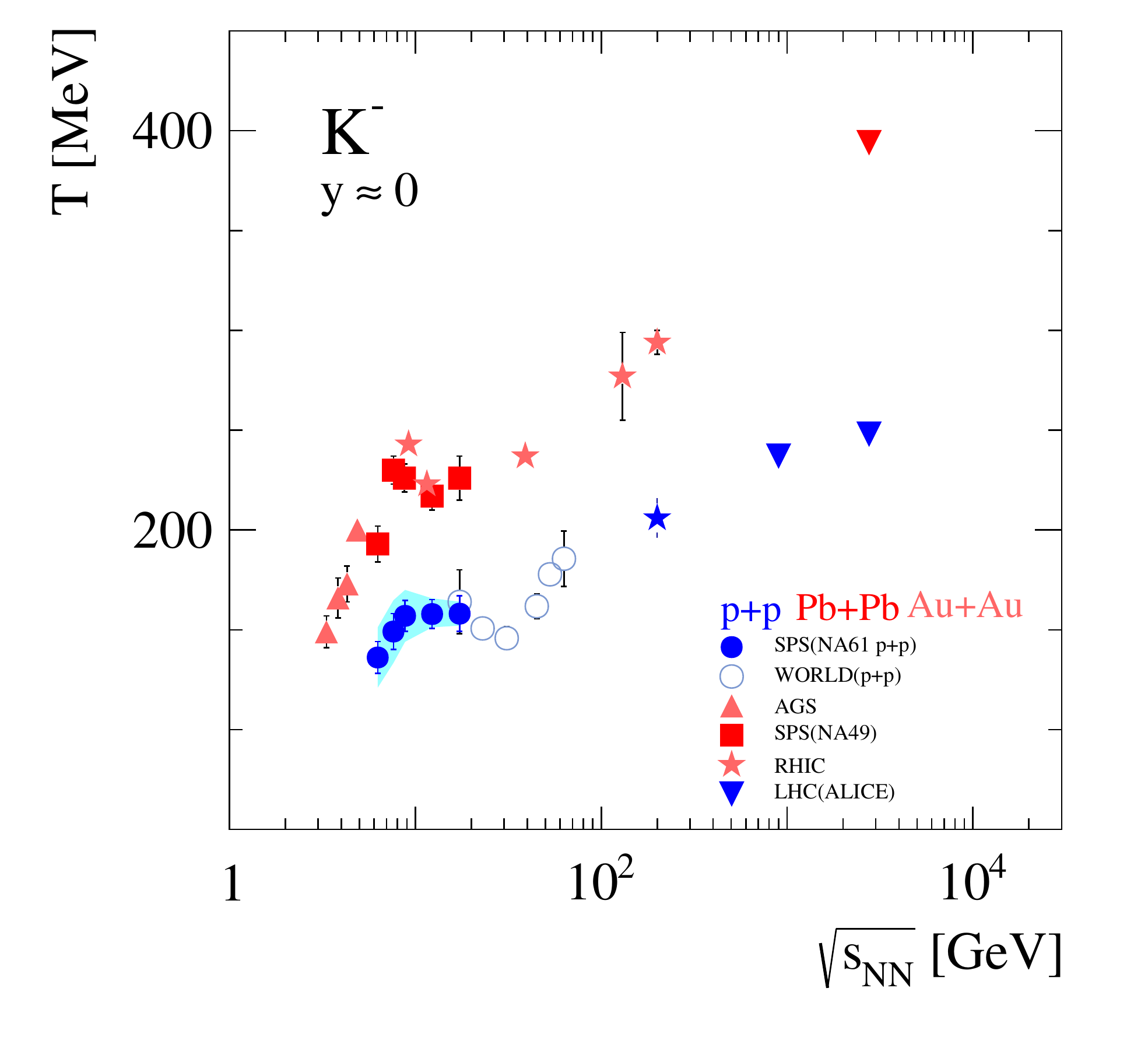}
	\includegraphics[width=0.48\textwidth]{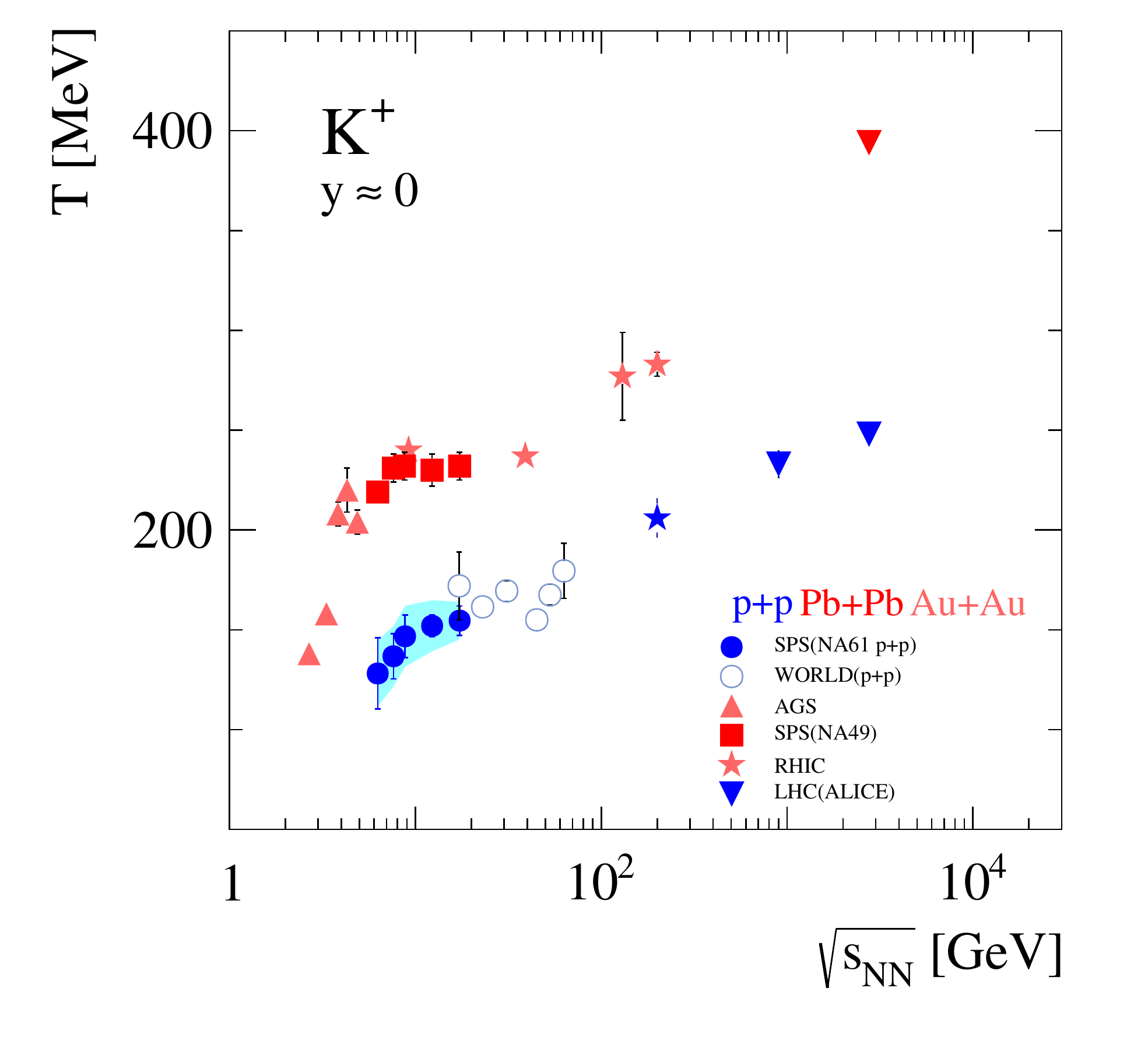}
	\end{center}
	\caption{Energy dependence of the inverse slope parameter $T$ of transverse mass spectra of kaons for inelastic p+p interactions and central Pb+Pb/Au+Au collisions.}
	\label{fig:step}
\end{figure}

Figure~\ref{fig:step} presents the energy dependence of the inverse slope parameter $T$ of kaons. Surprisingly the NA61/SHINE results from
inelastic p+p collisions exhibit rapid changes like observed in central Pb+Pb interactions. World data for p+p and Pb+Pb/Au+Au 
reactions are plotted for comparison and were taken from Refs.~\cite{Kliemant:2003sa,Abelev:2008ab,Abelev:2014laa,Aamodt:2011zj}.

The acceptance of the identification methods used at the moment by NA61/SHINE does not allow to measure spectra
of positive pions close to mid-rapidity.
Negatively charged pion spectra were measured  by NA61/SHINE~\cite{Abgrall:2013qoa}
using the $h^-$ method.
This method cannot be used for positively charged pions due to the large contribution of
protons and positive kaons.

\begin{figure}[!hc]
	\begin{center}
	\includegraphics[width=0.44\textwidth,valign=c]{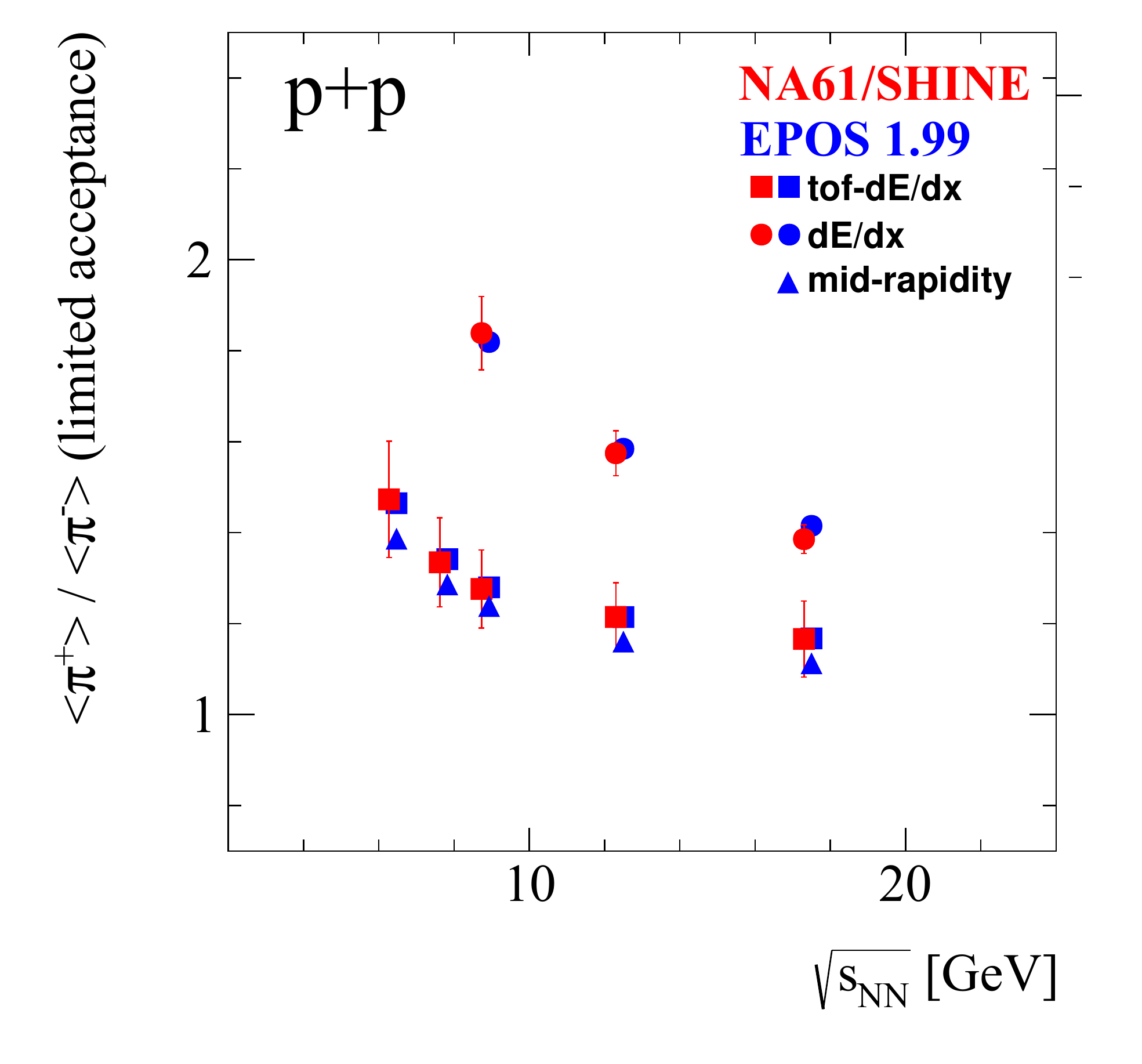}
	\includegraphics[width=0.44\textwidth,valign=c]{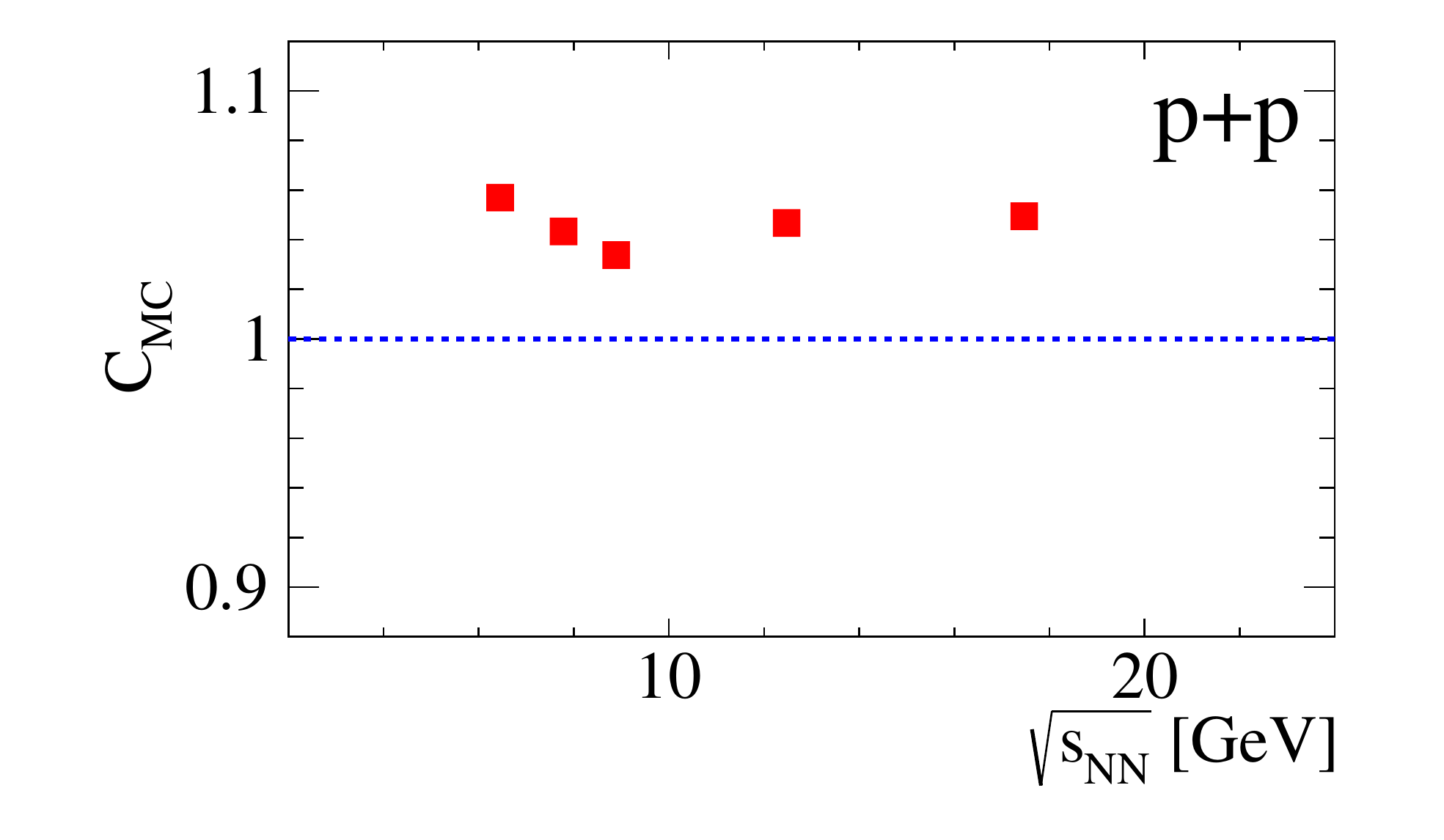}
	\end{center}
	\caption{The ratio of $\pi^{+}$ and $\pi^{-}$  yields  in inelastic p+p interactions as a function of collision energy.
The ratio was calculated in the acceptance of the $tof-dE/dx$ and $dE/dx$ identification methods. The NA61/SHINE results are compared to predictions of the EPOS model (left). Correction factor used to obtain $\pi^{+}$ yields from the measured $\pi^{-}$ yields (right), see Eq.~\protect\ref{eq:CMC}. }
	\label{fig:acceptance}
\end{figure}

Therefore, in order to determine the mid-rapidity yield of positively charged pions
the following procedure was used:
\begin{itemize}

\item
The ratio of measured $\pi^+$ and $\pi^-$ yields was calculated (within the acceptance
of the $tof-dE/dx$ and $dE/dx$ methods) and compared with model predictions as a function of collision energy, see Fig.~\ref{fig:acceptance}.
The agreement between the data and the EPOS model~\cite{Pierog:2009zt} predictions is better than 0.1\%.

\item
The mid-rapidity ratio of the $\pi^+$ and $\pi^-$ yields was calculated from the EPOS model.

\item The $\pi^+$ mid-rapidity yield was calculated as the product of the
measured $\pi^-$ yield at $y=0$, the measured $\frac{\pi+}{\pi^{-}}$ ratio in the $tof-dE/dx$ acceptance and the EPOS correction factor:
\begin{equation}
\pi^{+}\left(y=0\right)=\pi^{-}\left(y=0\right) \frac{\pi^{+}}{\pi^{-}}\left(tof-dE/dx\right)C_{MC},
\end{equation}
where:
\begin{equation}
C_{MC}=\left[\frac{\frac{\pi^{+}}{\pi^{-}}(y=0)} {\frac{\pi^{+}}{\pi^{-}}\left(tof\right)}\right]_{MC}\approx 5\%.
\label{eq:CMC}
\end{equation}
The final correction factor as a function of energy is shown in Fig.~\ref{fig:acceptance}.
\end{itemize}

\begin{figure}[!ht]
	\begin{center}
	\includegraphics[width=0.4\textwidth]{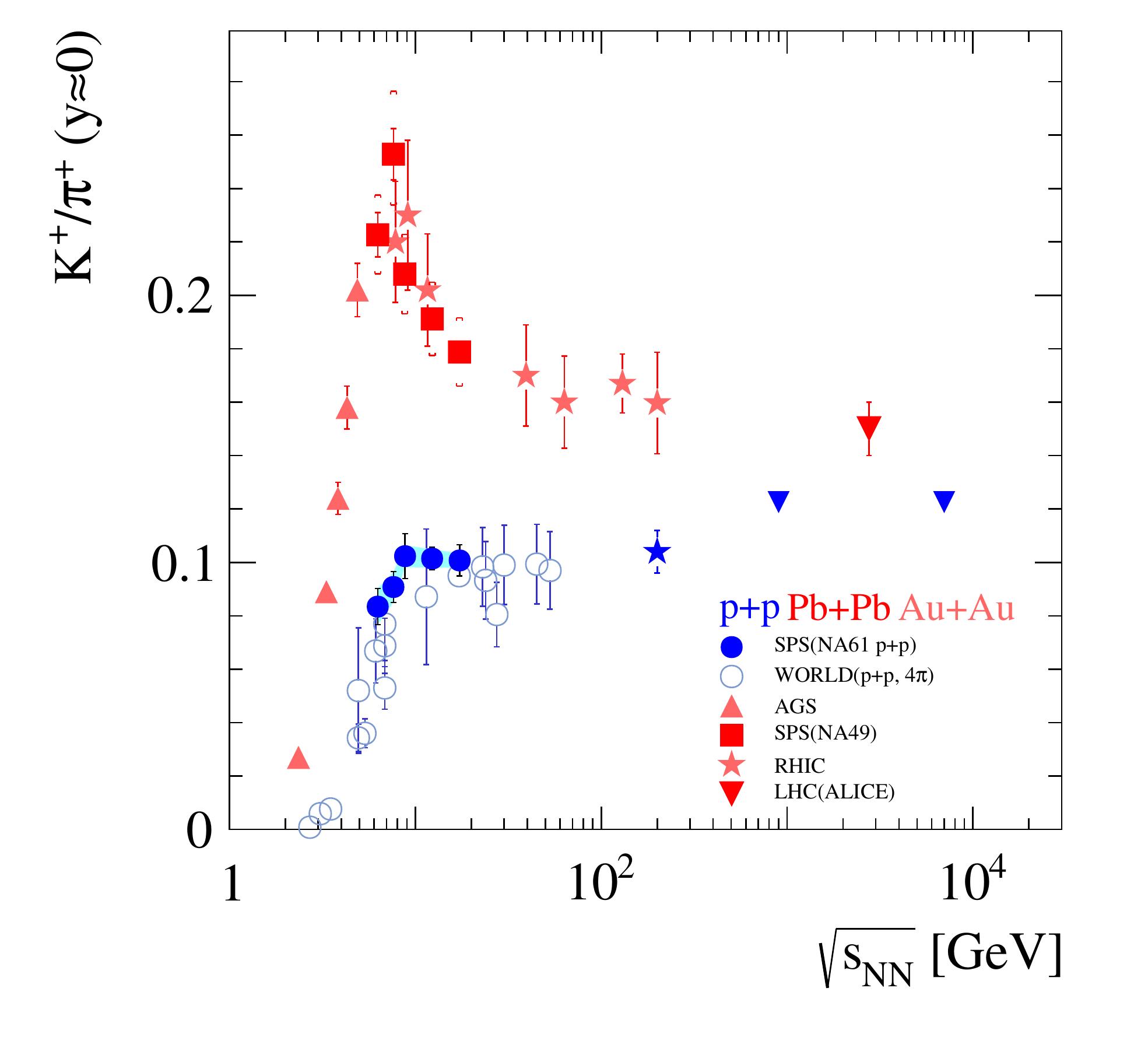}
	\includegraphics[width=0.4\textwidth]{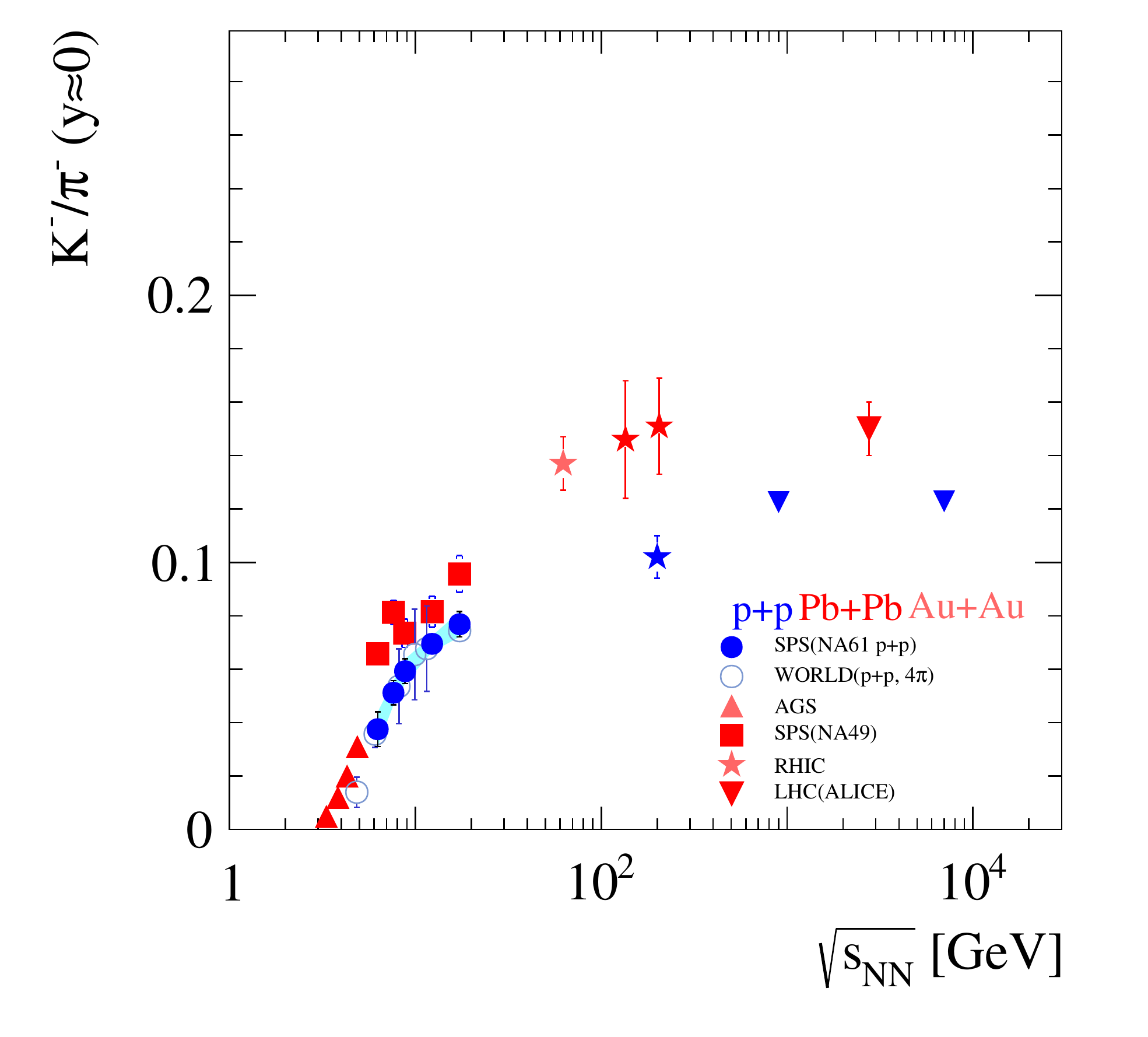}
	\end{center}
	\caption{The $K^{+}/\pi^{+}$ and $K^{-}/\pi^{-}$ ratios in inelastic p+p and central Pb+Pb/Au+Au reactions plotted versus collision energy.}
	\label{fig:horn}
\end{figure}

\begin{figure}[!ht]
	\begin{center}
	\includegraphics[width=0.4\textwidth]{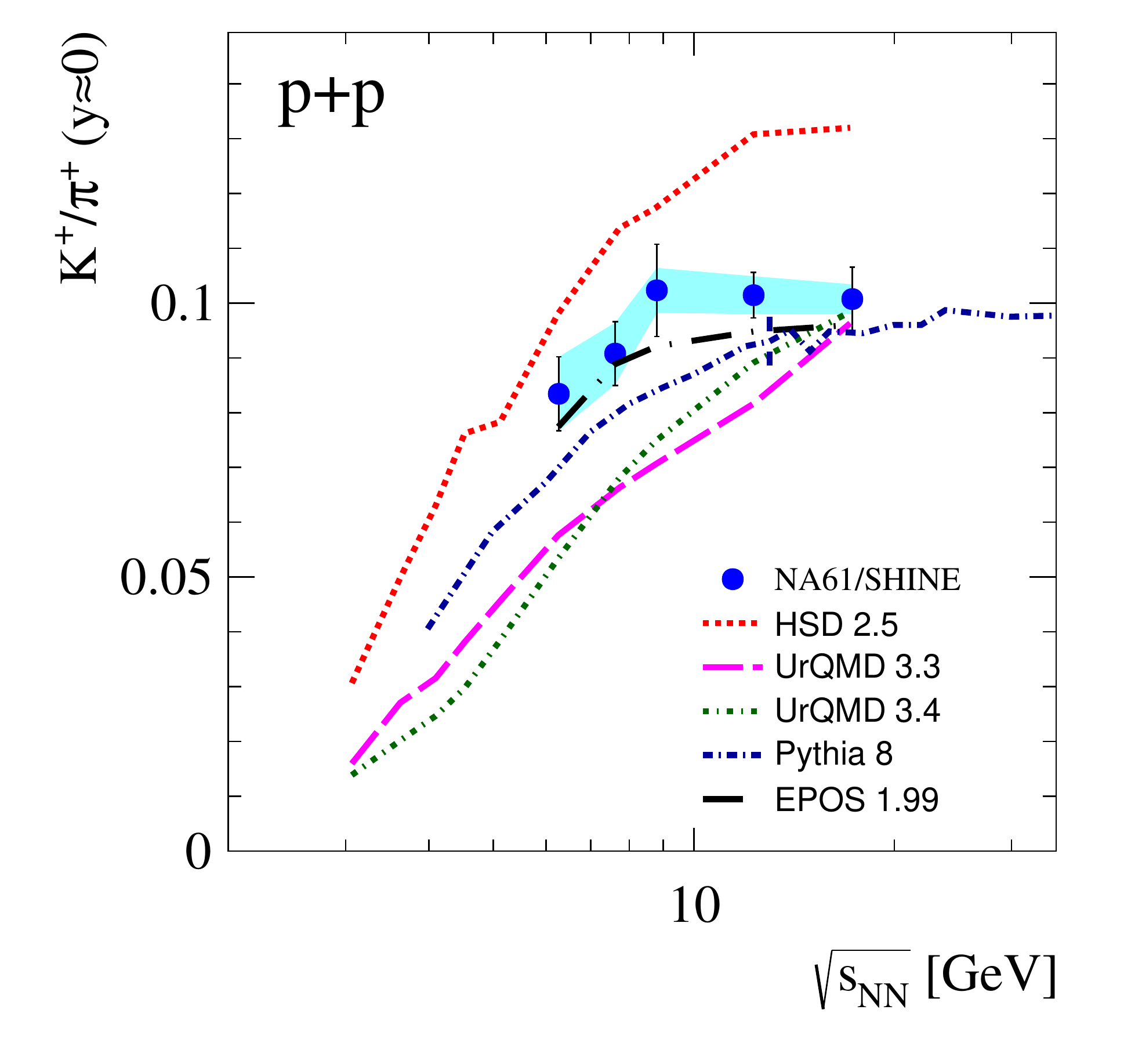}
	\includegraphics[width=0.4\textwidth]{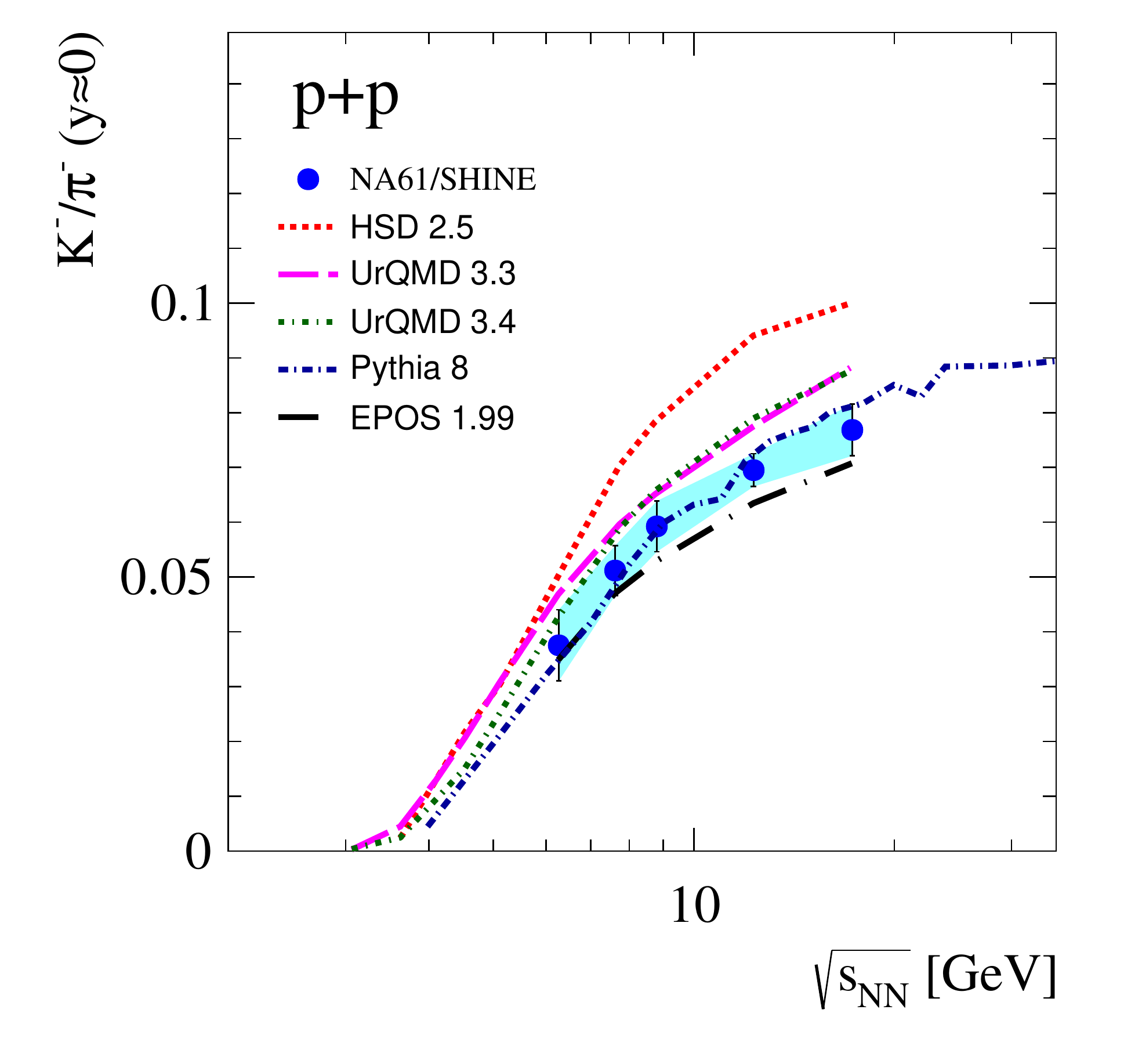}
	\end{center}
	\caption{Comparison of $K/\pi$ ratios in inelastic p+p collisions at SPS energies with theoretical models.}
	\label{fig:hornModels}
\end{figure}

The energy dependence of the $K^+ / \pi^+ $ and $K^- / \pi^- $ ratios at mid-rapidity for inelastic p+p interactions and central Pb+Pb/Au+Au collisions is presented in Fig.~\ref{fig:horn}. Surprisingly the NA61/SHINE data suggest that even in inelastic
p+p interactions the energy dependence of the $K^+ / \pi^+$ ratio exhibits rapid changes in the SPS energy range. However, the horn structure is significantly reduced/modified in comparison to that observed in central Pb+Pb collisions. Additionally world data~\cite{Gazdzicki:1995zs,Gazdzicki:1996pk,Arsene:2005mr,Aamodt:2011zj,Abelev:2012wca} is plotted to establish the trend outside the SPS energy range. The NA61/SHINE results were compared to theoretical models in Ref.~\cite{Vovchenko:2014ssa}, namely EPOS~\cite{Pierog:2009zt}, UrQMD~\cite{Bleicher:1999xi,Bass:1998ca}, Pythia 8~\cite{Sjostrand:2014zea} and HSD~\cite{Ehehalt:1996uq}. Figure~\ref{fig:hornModels} shows that these models do not describe well the NA61/SHINE results on p+p interactions. The high precision of the measurements should allow for significant improvement of the models.

\begin{figure}[ht]
	\begin{center}
	\includegraphics[width=0.4\textwidth,valign=c]{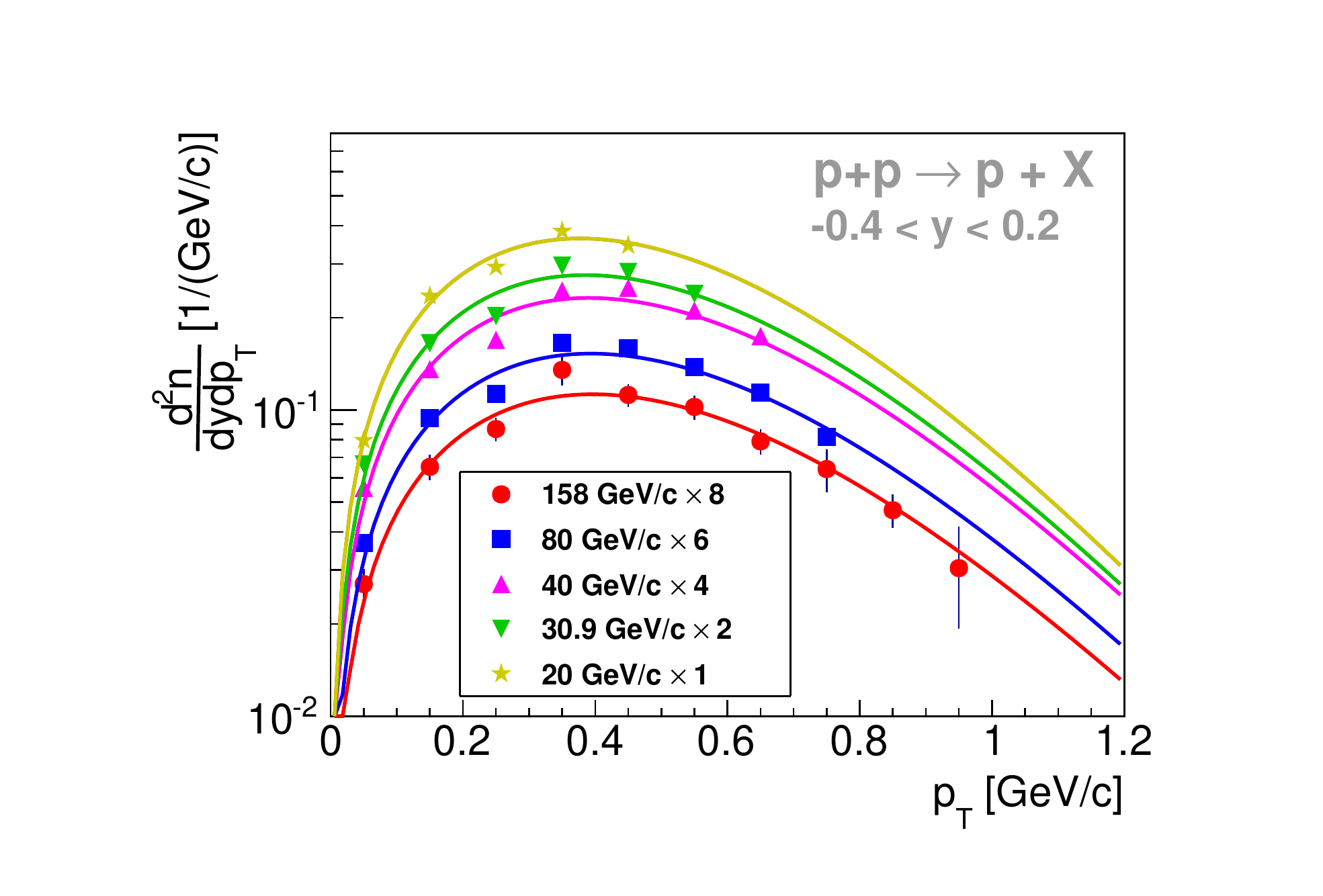}
	\includegraphics[width=0.4\textwidth,valign=c]{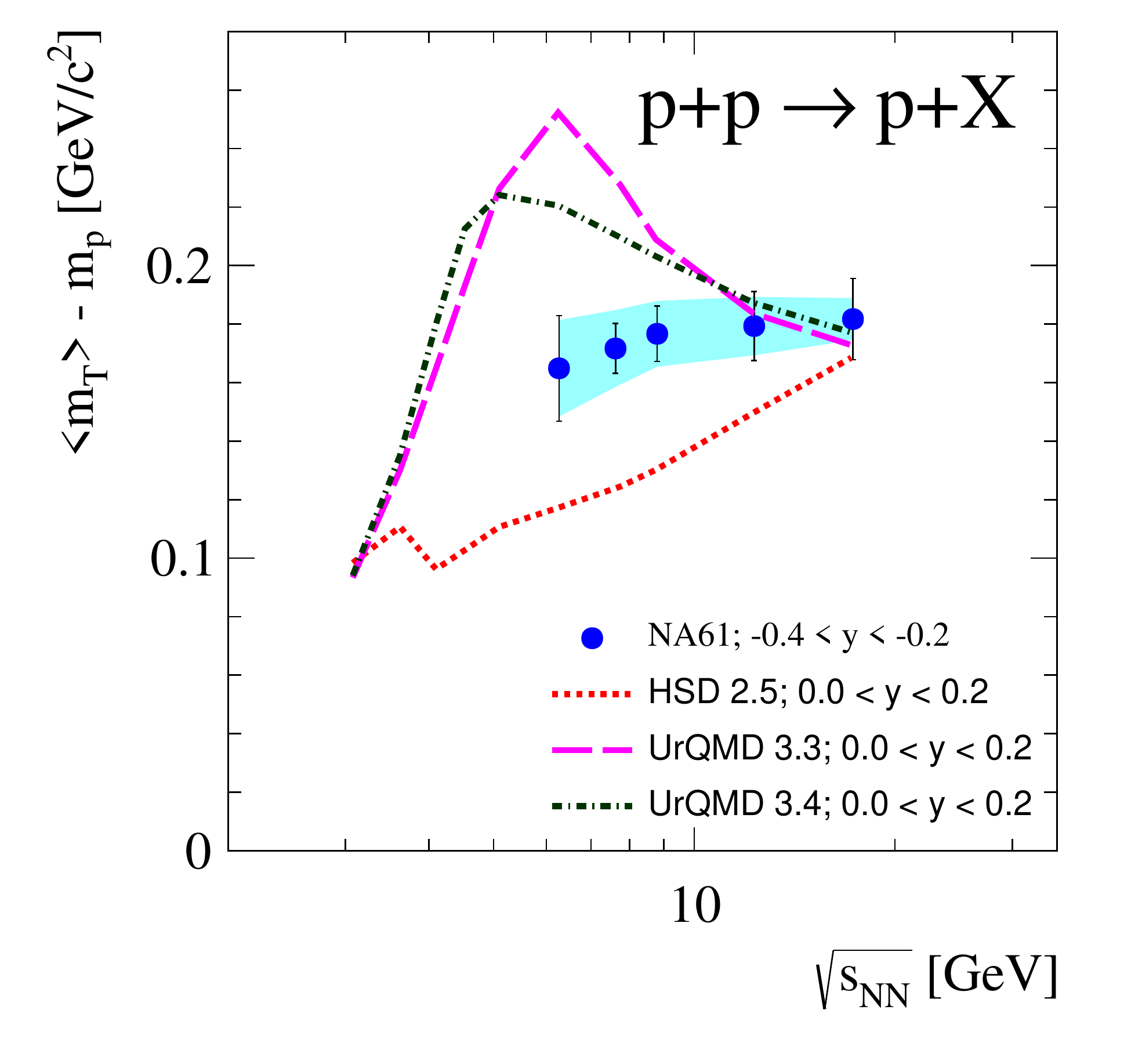}
	\end{center}
	\caption{Proton transverse momentum spectra (left) and mean transverse mass $m_{T}$ (right) around mid-rapidity in inelastic p+p interactions at SPS energies compared with theoretical models.}
	\label{fig:protons}
\end{figure}

Proton transverse momentum spectra and mean transverse mass $\langle m_{T} \rangle$ around mid-rapidity in inelastic p+p interactions at SPS energies are presented and compared with theoretical models in Fig.~\ref{fig:protons}.  Mean $m_{T}$ of protons was calculated from the data parametrization:
\begin{equation}
\frac{d^{2}n}{dp_{T}dy}=\frac{S~p_{T}}{T^{2}+m_{p}T}\exp\left(-\frac{\sqrt{p^{2}_{T}+
m^{2}_{p}}-m_{p}}{T}\right).
\end{equation}
It increases slowly with collision energy. Neither UrQMD nor HSD describe this behavior~\cite{Vovchenko:2014ssa}.

\section{Conclusions}
The NA61/SHINE measurements show that p+p collisions are unexpectedly interesting. Surprisingly the NA61/SHINE results suggest that even in inelastic p+p interactions the energy dependence of the $K^+ / \pi^+$ ratio and the inverse slope parameter of kaon transverse mass spectra exhibits rapid changes in the SPS energy range. However, the structures, e.g. the step and horn, are significantly reduced/modified compared to those observed in central Pb+Pb collisions. High precision NA61/SHINE data present challenges for models and should allow their improvement.

\section{Acknowledgments}
This work was supported by the National Science Center of Poland (grant UMO-2014/12/T/ST2/00692, grant UMO-2013/11/N/ST2/03879, grant UMO-2012/04/M/ST2/00816).


\begin{thebibliography}{99}
\bibitem{Abgrall:2013qoa}
N.~Abgrall et~al.
\newblock {Measurement of negatively charged pion spectra in inelastic p+p
  interactions at $p_{lab}$~=~20, 31, 40, 80 and 158 GeV/c}.
\newblock {\em Eur.Phys.J.}, C74(3):2794, 2014.

\bibitem{Pierog:2009zt}
T.~Pierog and K.~Werner.
\newblock {EPOS Model and Ultra High Energy Cosmic Rays}.
\newblock {\em Nucl.Phys.Proc.Suppl.}, 196:102--105, 2009.

\bibitem{Alt:2006dk}
C.~Alt et~al.
\newblock {Energy and centrality dependence of anti-p and p production and the
  anti-Lambda/anti-p ratio in Pb+Pb collisions between 20/A-GeV and 158/A-Gev}.
\newblock {\em Phys.Rev.}, C73:044910, 2006.

\bibitem{Alt:2007aa}
C.~Alt et~al.
\newblock {Pion and kaon production in central Pb + Pb collisions at 20-A and
  30-A-GeV: Evidence for the onset of deconfinement}.
\newblock {\em Phys.Rev.}, C77:024903, 2008.


\bibitem{Gazdzicki:2010iv}
Marek Gazdzicki, Mark Gorenstein, and Peter Seyboth.
\newblock {Onset of deconfinement in nucleus-nucleus collisions: Review for
  pedestrians and experts}.
\newblock {\em Acta Phys.Polon.}, B42:307--351, 2011.

\bibitem{Afanasiev:2002mx}
S.V. Afanasiev et~al.
\newblock {Energy dependence of pion and kaon production in central Pb + Pb
  collisions}.
\newblock {\em Phys.Rev.}, C66:054902, 2002.

\bibitem{Schnedermann:1993ws}
Ekkard Schnedermann, Josef Sollfrank, and Ulrich~W. Heinz.
\newblock {Thermal phenomenology of hadrons from 200-A/GeV S+S collisions}.
\newblock {\em Phys.Rev.}, C48:2462--2475, 1993.

\bibitem{Landau:1953gs}
L.D. Landau.
\newblock {On the multiparticle production in high-energy collisions}.
\newblock {\em Izv.Akad.Nauk Ser.Fiz.}, 17:51--64, 1953.

\bibitem{Shuryak:1974zc}
Edward~V. Shuryak.
\newblock {High-Energy Multiple Production of Hadrons and Landau Hydrodynamical
  Theory}.
\newblock 1974.

\bibitem{Klay:2003zf}
J.L. Klay et~al.
\newblock {Charged pion production in 2 to 8 agev central au+au collisions}.
\newblock {\em Phys.Rev.}, C68:054905, 2003.

\bibitem{Abbas:2013bpa}
Ehab Abbas et~al.
\newblock {Centrality dependence of the pseudorapidity density distribution for
  charged particles in Pb-Pb collisions at $\sqrt{s_{\rm NN}}$ = 2.76 TeV}.
\newblock {\em Phys.Lett.}, B726:610--622, 2013.

\bibitem{Adler:2003cb}
S.S. Adler et~al.
\newblock {Identified charged particle spectra and yields in Au+Au collisions
  at S(NN)**1/2 = 200-GeV}.
\newblock {\em Phys.Rev.}, C69:034909, 2004.

\bibitem{Abelev:2012wca}
Betty Abelev et~al.
\newblock {Pion, Kaon, and Proton Production in Central Pb--Pb Collisions at
  $\sqrt{s_{NN}} = 2.76$ TeV}.
\newblock {\em Phys.Rev.Lett.}, 109:252301, 2012.

\bibitem{Kliemant:2003sa}
Michael Kliemant, Benjamin Lungwitz, and Marek Gazdzicki.
\newblock {Energy dependence of transverse mass spectra of kaons produced in p
  + p and p + anti-p interactions: A Compilation}.
\newblock {\em Phys.Rev.}, C69:044903, 2004.

\bibitem{Gazdzicki:1995zs}
M.~Gazdzicki and D.~Roehrich.
\newblock {Pion multiplicity in nuclear collisions}.
\newblock {\em Z.Phys.}, C65:215--223, 1995.

\bibitem{Gazdzicki:1996pk}
Marek Gazdzicki and Dieter Rohrich.
\newblock {Strangeness in nuclear collisions}.
\newblock {\em Z.Phys.}, C71:55--64, 1996.

\bibitem{Abelev:2008ab}
B.I. Abelev et~al.
\newblock {Systematic Measurements of Identified Particle Spectra in $p p, d^+$
  Au and Au+Au Collisions from STAR}.
\newblock {\em Phys.Rev.}, C79:034909, 2009.

\bibitem{Abelev:2014laa}
Betty~Bezverkhny Abelev et~al.
\newblock {Production of charged pions, kaons and protons at large transverse
  momenta in pp and Pb????????Pb collisions at $\sqrt{s_{\rm NN}}$ =2.76 TeV}.
\newblock {\em Phys.Lett.}, B736:196--207, 2014

\bibitem{Aamodt:2011zj}
K.~Aamodt et~al.
\newblock {Production of pions, kaons and protons in $pp$ collisions at
  $\sqrt{s}= 900$ GeV with ALICE at the LHC}.
\newblock {\em Eur.Phys.J.}, C71:1655, 2011.

\bibitem{Arsene:2005mr}
I.~Arsene et~al.
\newblock {Centrality dependent particle production at y=0 and y ~ 1 in Au + Au
  collisions at s(NN)**(1/2) = 200-GeV}.
\newblock {\em Phys.Rev.}, C72:014908, 2005.

\bibitem{Bass:1998ca}
S.A. Bass, M.~Belkacem, M.~Bleicher, M.~Brandstetter, L.~Bravina, et~al.
\newblock {Microscopic models for ultrarelativistic heavy ion collisions}.
\newblock {\em Prog.Part.Nucl.Phys.}, 41:255--369, 1998.

\bibitem{Bleicher:1999xi}
M.~Bleicher, E.~Zabrodin, C.~Spieles, S.A. Bass, C.~Ernst, et~al.
\newblock {Relativistic hadron hadron collisions in the ultrarelativistic
  quantum molecular dynamics model}.
\newblock {\em J.Phys.}, G25:1859--1896, 1999.

\bibitem{Ehehalt:1996uq}
W.~Ehehalt and W.~Cassing.
\newblock {Relativistic transport approach for nucleus nucleus collisions from
  SIS to SPS energies}.
\newblock {\em Nucl.Phys.}, A602:449--486, 1996.

\bibitem{Sjostrand:2014zea}
Torbjorn Sjostrand, Stefan Ask, Jesper~R. Christiansen, Richard Corke,
  Nishita Desai, et~al.
\newblock {An Introduction to PYTHIA 8.2}.
\newblock 2014.

\bibitem{Vovchenko:2014ssa}
V.~Yu. Vovchenko, D.V. Anchishkin, and M.I. Gorenstein.
\newblock {System-size and energy dependence of particle momentum spectra: The
  UrQMD analysis of $p+p$ and Pb+Pb collisions}.
\newblock {\em Phys.Rev.}, C90(2):024916, 2014.

\end{thebibliography}
\end{document}